\documentclass[12pt,aps,prd,preprint,tightenlines,
  showkeys,showpacs,nofootinbib]{revtex4}

\newcommand{\gev}{\text{GeV}}

\newcommand{\cm}{\text{cm}}

\newcommand{\s}{\text{s}}

\newcommand{\eqref}[1]{Eq.~(\ref{#1})}

\newcommand{\figref}[1]{Fig.~\ref{fig:#1}}

\newcommand{\rem}[1]{{\bf #1}}
\usepackage{hyperref}      
\usepackage{epsfig}

\preprint{UCI-TR-2012-18}
\newcommand{\text}[1]{{\rm #1}}
\newcommand{\implies}{\Rightarrow}

\begin{document}

\title{\vspace*{2in}
Dark Matter and Indirect Detection in Cosmic Rays
\rem{\footnote{To appear in the Proceedings of {\em
      Centenary Symposium 2012: Discovery of Cosmic Rays}, Denver,
    Colorado, June 2012.}}
\vspace*{0.7in}
}




\author{Jonathan L.~Feng}
\affiliation{Department of Physics and Astronomy, University of
California, Irvine, CA 92697, USA
\vspace*{.5in}
}

\begin{abstract}
\vspace*{.2in}
In the early years, cosmic rays contributed essentially to particle
physics through the discovery of new particles.  Will history repeat
itself?  As with the discovery of the charged pion, the recent
discovery of a Higgs-like boson may portend a rich new set of
particles within reach of current and near future experiments. These
may be discovered and studied by cosmic rays through the indirect
detection of dark matter.
\vspace*{.5in}
\end{abstract}

\maketitle



\section{Will History Repeat Itself?}

At this conference, we are celebrating 100 years of cosmic rays and
looking to the future.  As has been recounted here, the early years of
cosmic rays were a glorious period, in part because cosmic rays
contributed to the birth of particle physics through the discovery of
new particles, including the positron, the muon, and the pion.

These discoveries were in some cases serendipitous, but let me focus
on one that was not --- the discovery of the charged pion, whose
existence was predicted well in advance of its discovery.  The history
surrounding this discovery is well-known and fascinating and may be
very briefly summarized as follows:
\begin{itemize}
\item 1935: To explain the strong nuclear force, Yukawa postulates new
  particle physics at the 100 MeV mass scale~\cite{Yukawa:1935xg}.
\item 1947: A boson is discovered in this mass range, associated with
  a broken (global) symmetry: the charged pion~\cite{Lattes:1947mw}.
\item The next 20 years: Many accompanying particles are discovered
  and studied by both cosmic rays and particle accelerators.
\end{itemize}

Is a similar story developing in physics today?  We have just
witnessed another discovery of extraordinary importance, which may
play out with striking similarities:
\begin{itemize}
\item 1934: To explain the weak nuclear force, Fermi postulates new
  particle physics at the 100 GeV mass scale~\cite{Fermi:1934hr}.
\item 2012: A boson is discovered in this mass range, associated with
  a broken (gauge) symmetry: the Higgs
  boson~\cite{ATLAS:2012gk,CMS:2012gu}.
\item The next 20 years: Many accompanying particles are discovered
  and studied in both cosmic rays and particle accelerators.
\end{itemize}

Of course, the last point is still somewhat uncertain (for nitpickers,
even the penultimate point requires confirmation), but there are good
reasons to expect additional new particles at the weak scale.  The
Higgs boson mass is highly fine-tuned.  All attempts to explain this
fine-tuning predict new particles at the weak scale.  This motivation
may be viewed as an aesthetic one, but it is buttressed by another:
the need for dark matter.  Although there are many dark matter
candidates, some particles with mass at the 100 GeV scale, the
so-called weakly-interacting massive particles (WIMPs), have a
privileged position.  If one assumes a new particle $X$ that was
initially in thermal equilibrium in the early Universe, its relic
density is determined by its annihilation cross section $\sigma_A$.
The relation is remarkably simple:
\begin{equation}
\Omega_X \propto \frac{1}{\langle \sigma_A v \rangle} \sim
\frac{m_X^2}{g_X^4} \ .
\label{relicdensity}
\end{equation}
The last expression is simply the result of dimensional analysis,
where $m_X$ is the dark matter's mass, and $g_X$ is the characteristic
coupling that enters the dominant annihilation processes.  If one
assumes $g_X \sim 1$ and includes the neglected dimensionless
parameters in \eqref{relicdensity}, one finds that
\begin{equation}
\Omega_X \sim 0.1 \implies m_X \sim 100~\gev \ ;
\end{equation}
that is, requiring the new particle to have the right relic density to
be dark matter requires its mass to be near the weak scale.  This
remarkable coincidence, the ``WIMP miracle,'' implies that particle
physics and cosmology independently point to the weak scale as a
promising place to look for new particles.

\section{WIMP Dark Matter Detection}

\begin{figure}
  \includegraphics[height=0.29\textheight]{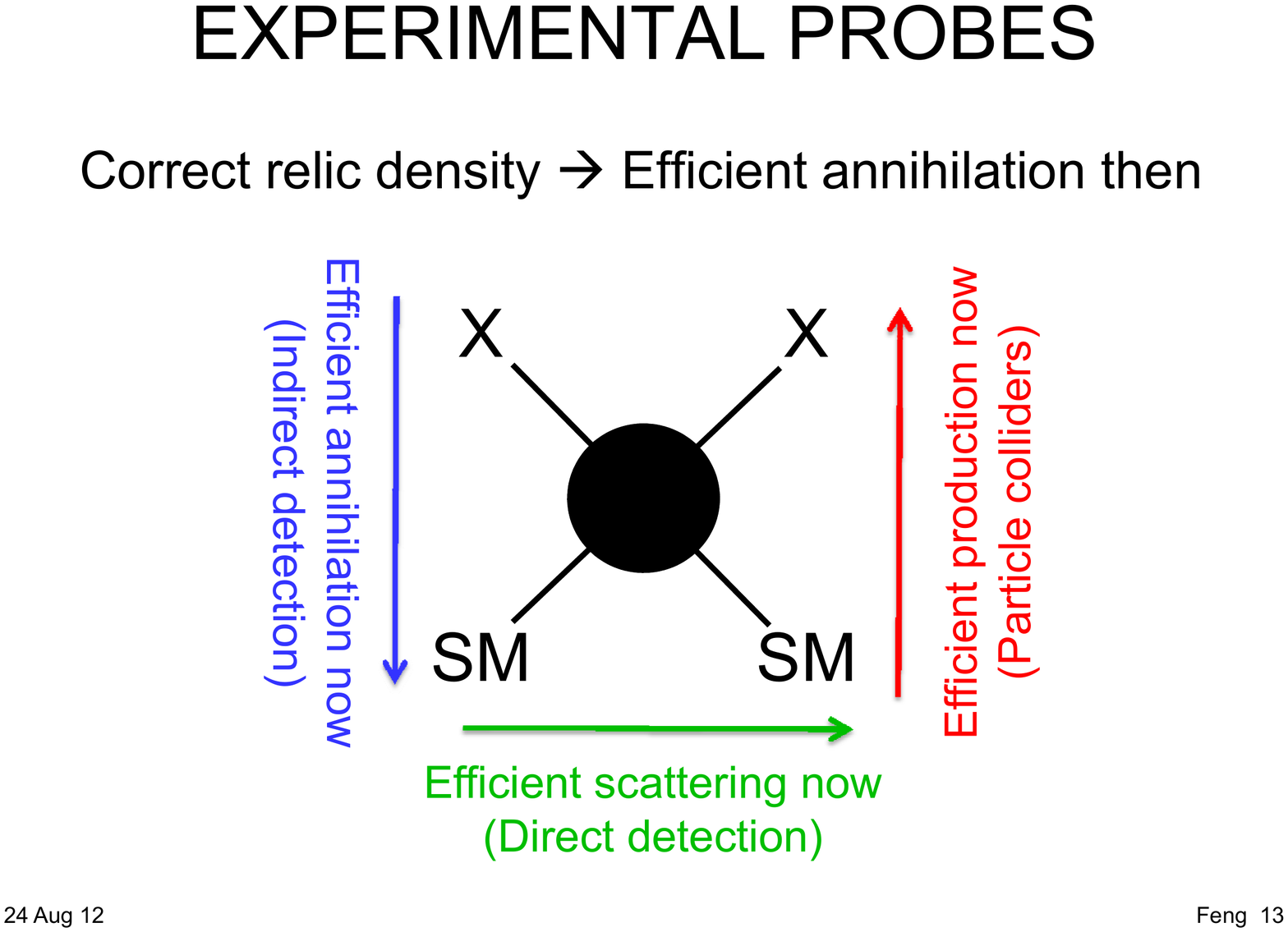}
  \caption{Dark Matter complementarity: For WIMPs to have the right
    relic density to be dark matter, they must annihilate efficiently
    in the early Universe.  This implies efficient annihilation now
    and signals for indirect detection experiments, efficient
    scattering now and signals for direct detection experiments, and
    efficient production now at particle colliders.}
\label{fig:complementarity}
\end{figure}

The WIMP miracle not only motivates a class of dark matter candidates,
it also tells us how to look for them.  As illustrated in
\figref{complementarity}, the WIMP miracle requires efficient
annihilation in the early Universe.  Assuming annihilation is
dominantly to known particles, this implies a four-particle
$X$-$X$-SM-SM interaction, where SM denotes a standard model particle.
This in turn implies that dark matter can be discovered through
present day annihilation $X X \to \text{SM} \ \text{SM}$ (indirect
detection), through scattering $X \ \text{SM} \to X \ \text{SM}$
(direct detection), and by producing it at colliders through
$\text{SM} \ \text{SM} \to X X$ (provided the final state dark matter
particles are accompanied by some visible particles).

In fact, not only does the WIMP miracle tell us how to look for dark
matter, it also tells us (roughly) when to give up.  Although we know
little about dark matter, we do know that there cannot be too much of
it.  In the WIMP paradigm, this implies the four point interactions of
\figref{complementarity} cannot be too weak, providing a floor to the
most motivated annihilation, scattering, and production cross
sections.

Here we will focus on indirect detection.  The experimental program in
this field may be summarized as attempts to fill in the following
sentence in promising ways:

\begin{center}
WIMPs annihilate in $\langle \text{\it a\ place} \rangle$ to
  $\langle \text{\it particles} \rangle$ that are detected by $\langle
  \text{\it an\ experiment} \rangle$.
\end{center}

There are many ways to complete this sentence, and the indirect
detection of dark matter is an extremely diverse, active, and exciting
field.  In the following, I will outline just three of the many
possible directions being explored at present.

\section{Indirect Detection in Positrons}

\begin{figure}
  \includegraphics[height=0.29\textheight]{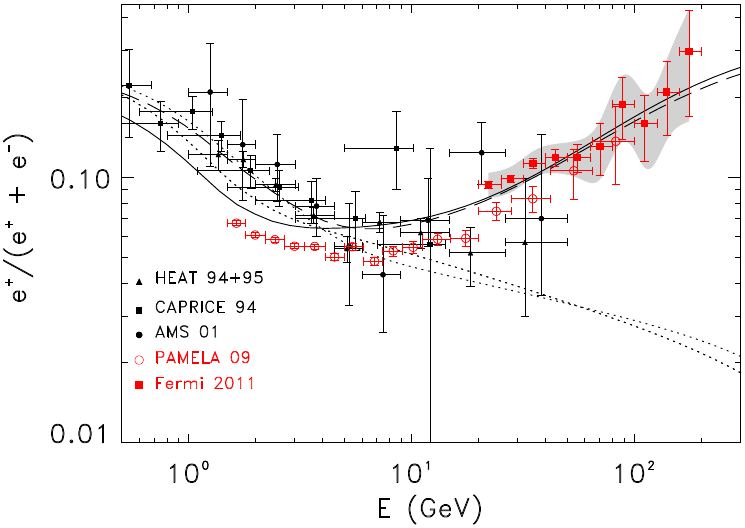}
  \caption{The positron fraction as a function of energy as recently
    measured by PAMELA~\cite{Adriani:2010ib} and
    Fermi-LAT~\cite{FermiLAT:2011ab} up to energies of $\sim
    100~\gev$.  The lower dotted curves are the expectation from
    standard astrophysical background~\cite{Moskalenko:1997gh}, and
    the higher solid and dashed curves include possible contributions
    from pulsars or dark matter annihilation~\cite{Grasso:2011wt}. }
\label{fig:positrons}
\end{figure}

WIMP dark matter may annihilate in the galactic halo to positrons that
are detected by satellites and balloon-borne experiments.  Following
earlier measurements from HEAT~\cite{Barwick:1997ig} and
AMS-01~\cite{Aguilar:2007yf} of the positron spectrum up to energies
of $\sim 10~\gev$, this field has been re-animated with results from
PAMELA~\cite{Adriani:2010ib} up to energies $\sim 100~\gev$ that show
positron fractions that increase with energy.  This result has been
confirmed recently by a clever analysis from
Fermi-LAT~\cite{FermiLAT:2011ab} (See \figref{positrons}.)  This
increase is in conflict with standard expectations for astrophysical
background~\cite{Moskalenko:1997gh}, leading some to explore the
possibility that the excess is a signal of dark matter annihilation.

Unfortunately, the signal is far larger than expected in the WIMP
paradigm.  As noted above, the requirement that WIMPs have the correct
thermal relic density implies a characteristic annihilation cross
section.  The annihilation cross section required to reproduce the
signal is 100 to 1000 times bigger, and so requires enhancements from
particle physics that exploit the different kinematics of dark matter
annihilation in the early universe and now.  Alternatively, one may
sacrifice the WIMP miracle, and dark matter may have a completely
different production mechanism.  In the meantime, researchers have
recalled that pulsars may enhance the positron fraction with excesses
that are of the required size~\cite{1987ICRC....2...92H,%
  1989ApJ...342..807B,1995A&A...294L..41A,2001A&A...368.1063Z}.  At
present, pulsars appear to be by far the more natural and conservative
explanation. Further progress awaits data from AMS-02 and proposed
experiments, such as CALET.

\section{Indirect Detection in Neutrinos}

WIMP dark matter may also annihilate in the center of the Sun to
neutrinos that are detected by neutrino telescopes.  A WIMP is
captured by the Sun when it scatters off normal matter in the Sun and
its velocity is reduced below escape velocity.  This process implies
that WIMPs build up in the Sun, providing a relatively large and
nearby overdensity that enhances the annihilation signal.

For most WIMP candidates, the WIMP population in the Sun has reached
equilibrium.  The annihilation and capture rates are therefore equal,
implying a relation between the annihilation and scattering cross
sections.  Indirect detection bounds may therefore be compared to
direct detection bounds.

\begin{figure}
  \includegraphics[height=0.29\textheight]{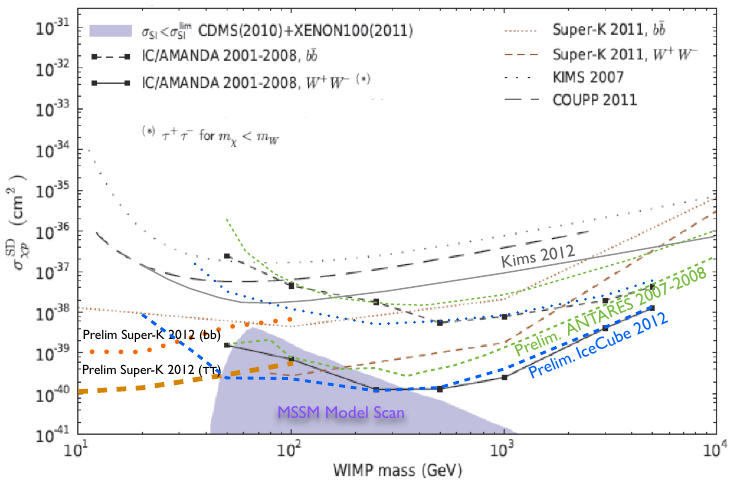}
  \caption{Limits on the WIMP-proton spin-dependent scattering cross
    section from searches for WIMPs annihilating to neutrinos in the
    Sun from the neutrino telescopes
    SuperKamiokande~\cite{Tanaka:2011uf},
    ANTARES~\cite{Bertin:2012fb}, and IceCube~\cite{Rott:2012gh}.
    These indirect search limits assume that annihilation is
    dominantly to bottom quarks, $\tau$ leptons, and $W$ bosons, as
    indicated, and are compared to the direct search limits from the
    KIMS and COUPP experiments.}
\label{fig:neutrinos}
\end{figure}

Preliminary bounds on spin-dependent WIMP-proton scattering from
SuperKamiokande~\cite{Tanaka:2011uf}, ANTARES~\cite{Bertin:2012fb},
and IceCube~\cite{Rott:2012gh} are shown in \figref{neutrinos}.  These
limits are compared to direct detection bounds, and also to the
expectations for neutralino dark
matter~\cite{Goldberg:1983nd,Ellis:1983ew}. The indirect searches are
seen to be the most stringent at present, and are probing the
parameter space of well-motivated WIMP models.  Remarkably, spin-{\em
  independent} probes from indirect detection are also becoming
competitive with direct detection.  The prospects for improved
sensitivity are excellent as experiments continue to gather data, and
new developments, such as the proposed in-fill array PINGU at IceCube,
may greatly improve the experimental sensitivity to dark matter.

\section{Indirect Detection in Gamma Rays}

Last, WIMPs may annihilate in the galactic center or in dwarf galaxies
to photons that are detected by space-based and balloon-borne
experiments or by ground-based atmospheric Cherenkov telescopes.
Traditionally the galactic center has been the favorite target for
such sources, given its large overdensity, but the recent discovery of
many dwarf galaxies has made them another prime target, with the lower
dark matter densities compensated by the promise of reduced and better
understood background.

\begin{figure}
  \includegraphics[height=0.29\textheight]{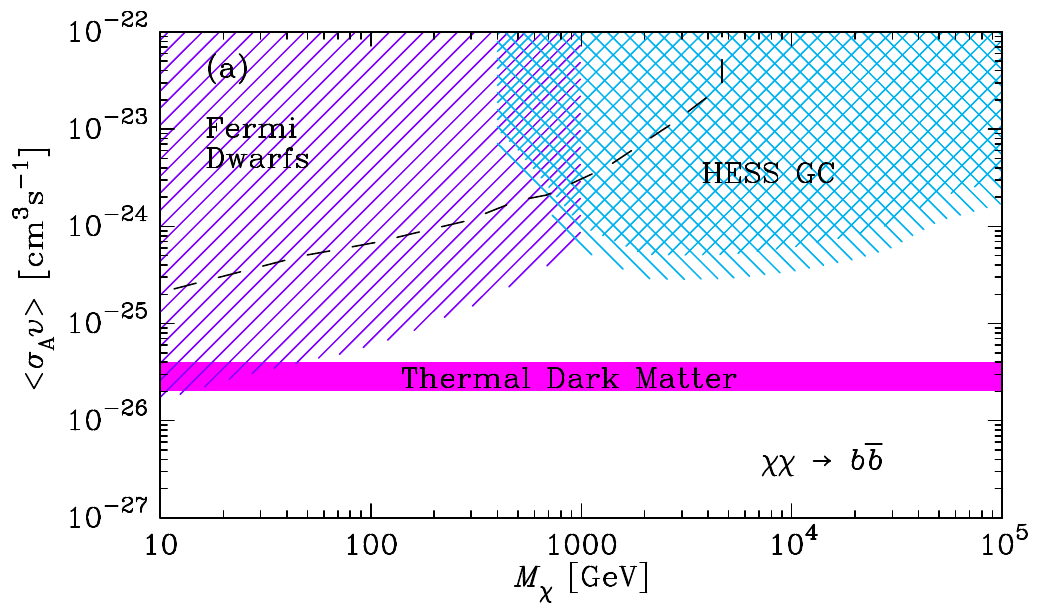}
  \caption{Bounds on dark matter annihilation cross sections from
    limits on continuum gamma ray fluxes from dwarf galaxies and the galactic
    center~\cite{Abazajian:2011ak}.
}
\label{fig:gammarays}
\end{figure}

Photon signals are of two kinds: continuum signals from dark matter
annihilation to other particles with a radiated photon, and line
signals from dark matter annihilating directly to $\gamma X$, where
most typically $X = \gamma, Z, h$.  The continuum signal has a smooth
energy distribution, but the expected continuum flux in most models is
typically far larger than the line signal.  Current bounds on dark
matter annihilation cross sections from
Fermi-LAT~\cite{GeringerSameth:2011iw,Ackermann:2011wa} and
HESS~\cite{Abramowski:2011hc} are given in \figref{gammarays}.  Also
shown for reference is the annihilation cross section $\langle
\sigma_A v \rangle \simeq 3 \times 10^{-26}~\cm^3/\s$, which is what
is required for dark matter to have the right relic density if it
annihilates through $s$-channel processes.  Some well-known dark
matter candidates, such as the Kaluza-Klein
photon~\cite{Servant:2002aq,Cheng:2002ej} in extra dimensional
theories are $s$-channel annihilators, but some, such as the
neutralino~\cite{Goldberg:1983nd,Ellis:1983ew} from supersymmetry, are
not and predict reduced values of $\langle \sigma_A v \rangle$
now. However, the fact that current bounds are approaching this
important reference value is a measure of the promise for these
experiments to probe viable thermal relic models.

The line signal is in principle much easier to distinguish from
backgrounds, but since dark matter does not couple directly to
photons, the line signal is typically expected to proceed only through
loops in the Feynman diagram, and so is typically highly suppressed.
At present, much activity and excitement surrounds a tentative line
signal at $E_{\gamma} \simeq 135~\gev$~\cite{Weniger:2012tx}.  The
required annihilation cross section to explain this signal is very
large, but models with such large signals existed even before the
anomaly was reported~\cite{Baltz:2002we,Jackson:2009kg}, and, of
course, many more have been constructed since.  Further progress to
determine if the line signal is real and to improve sensitivities for
both continuum and line searches is sure to come from continued
running of existing experiments and upcoming experiments, including
HESS-2, HAWC, CTA, DAMPE, GAMMA-400, HERD, as well as AMS-02 and
CALET.

\begin{acknowledgments}
I thank Jonathan Ormes for organizing a conference filled with both
fascinating history and science, and Carsten Rott for guiding me to
the latest neutrino results.  This work is supported in part by NSF
grant PHY-0970173 and in part by a Simons Fellowship in Theoretical
Physics.
\end{acknowledgments}




\providecommand{\href}[2]{#2}\begingroup\raggedright\endgroup

\end{document}